\documentclass[12pt]{article}
\RequirePackage[OT1]{fontenc}
\RequirePackage{amsthm,amsmath}
\usepackage{accents}
\usepackage{breqn}
\usepackage{soul}
\usepackage[numbers,sort&compress]{natbib}
\RequirePackage[citecolor=blue,urlcolor=blue]{hyperref}
\usepackage{soul,xcolor}
\usepackage{hyperref}
\usepackage{cleveref}
\usepackage{amsfonts}
\usepackage{amssymb}
\usepackage{graphicx}
\usepackage{graphics}
\usepackage{setspace}
\usepackage{physics}

\numberwithin{equation}{section}
\newcommand{\curly}{\mathrel{\leadsto}}

\def\b1{{\mathbf 1}}
\def\b0{{\mathbf 0}}

\def\bz{{\boldsymbol z}}

\def\cU{{\mathcal U}}

\def\cI{{\mathcal I}}

\def\cT{{\mathcal T}}

\def\bbR{{\mathbb R}}

\def\bbI{{\mathbb I}}

\begin{document}

\title{Lorentz covariant physical Brownian motion: Classical and quantum (Revised)\\
Final version to appear in AoP (\url{https://doi.org/10.1016/j.aop.2024.169857})}
\author{Henryk Gzyl\\
Centro de Finanzas IESA, Caracas, Venezuela.\\
 henryk.gzyl@iesa.edu.ve}
\date{}
 \maketitle

\setlength{\textwidth}{4in}

\vskip 1 truecm
\baselineskip=1.5 \baselineskip \setlength{\textwidth}{6in}

\begin{abstract}
In this work, we re-examine the Goldstein-Ka\c{c} (also called Poisson-Ka\c{c}) velocity switching model from two points of view. On the one hand, we prove that the forward and backward Chapman-Kolmogorov equations of the stochastic process are Lorentz covariant when the trajectories are parameterized by their proper time. On the other hand, to recast the model as a quantum random evolution, we restate the Goldstein-Ka\c{c} model as a Hamiltonian system, which can then be quantized using the standard correspondence rules. It turns out that the density matrix for the random quantum evolution satisfies a Chapman-Kolmogorov equation similar to that of the classical case, and therefore, it is also Lorentz covariant. To finish, we verify that the quantum model is also consistent with special relativity and that transitions outside the light cone, that is, transitions between states with disjoint supports in space-time, cannot occur. 
\end{abstract}

\noindent \textbf{Keywords}: Quantum systems subject to random pulses, Random evolutions, Brownian motion, Lorentz covariance of transport equations.\\

\begin{spacing}{0.50}
\small{\tableofcontents}
\end{spacing}

\section{Introduction and Preliminaries} 
The aim of this paper is to prove that, the classical Goldstein-Ka\c{c} velocity switching model, and its quantized version are Lorentz covariant. To put this objective in a better perspective, let us explain why the standard model for Brownian motion and the Sch\"ordinger approach to quantum mechanics are incompatible with special relativity. Consider the simple case of a homogeneous, isotropic diffusion, or that of a free quantum particle moving on the real axis. In the first case, the time evolution of the probability density is given by:

\begin{equation}\label{eq1.1}
\frac{\partial}{\partial t}\rho(t,x) =\frac{1}{2}\frac{\partial^2}{\partial x^2}\rho(t,x);\;\;\;\rho(0,x)=\rho_0(x).
\end{equation}
Here $\rho(t,x)$ stands for the probability density of finding the diffusing particle at point $x$ at time $t.$ The factor $1/2$ is standard convention in the mathematical literature. It amounts to have a standard deviation $\sigma=1.$

Similarly, if $\psi(t,x)$ denotes the probability amplitude of finding a particle (of mass $m=1$) at point $x$ at time $t,$ then it satisfies the (Schr\"odinger's) equation:

\begin{equation}\label{eq1.2}
i\hbar\frac{\partial}{\partial t}\psi(t,x) =\frac{-\hbar^2}{2}\frac{\partial^2}{\partial x^2}\psi(t,x);\;\;\;\psi(0,x)=\psi_0(x).
\end{equation}

If, as assumed in the special relativity theory, no particle can move with a speed larger than the speed of light $c,$ then these two equations can not be compatible with it, because both imply that particles can move with a speed larger than $c.$ To see it, suppose that the initially the particle is in a small interval $[-a,a]$ about the origin, that is, that both $\rho_0(x)$ and $|\psi_0(x)|^2$ vanish off $[-a,a].$ It is trivial to verify that for any interval $\cI,$ that does not intersect $[-a-ct,a+ct]$ we have:

\begin{equation}\label{incomp}
\int_{\cI}\rho(t,x)dx > 0,\;\;\;\mbox{and}\;\;\;\int_{\cI}|\psi(t,x)|^2dx > 0.
\end{equation}
This implies that particles can move from $[-a,a]$ to $\cI$ at speeds larger than $c.$ To phrase it in terms of the summation of probabilities over {\it all possible} trajectories joining a point $x_1$ to a point $x_2$ between any times $t_1$ and $t_2>t_1.$ Observe that, among these trajectories, there are uncountably many that wander away from the interval $[x_1,x_2]$ by distances larger that $c(t_2-t_1).$  To sum up, the underlying process described by either \eqref{eq1.1} or by \eqref{eq1.2} can move at speeds larger than the speed of light.\\
Moreover, a standard result in probability theory asserts that, with respect to the Wiener measure on the space of all continuous trajectories, the class of all everywhere non-differentiable curves has probability $1.$ In other words, physical, continuously differentiable trajectories have $0$ probability of occurrence. For this result in Brownian motion theory, see \cite{MP}. This fact is relevant, because the action along a classical trajectory involves the square of the derivative, and the action along classical trajectories plays an essential role in the path integral formulation of quantum mechanics.

An attempt to go around this issue, in the theory of heat propagation governed by (Fourier's equation) \eqref{eq1.1} is considered in \cite{Cat}. In that work,  an argument from kinetic theory is used to obtain a telegraph equation (like \eqref{syst2} below) governing the propagation of heat, instead of the Fourier's equation \eqref{eq1.1}. 

\subsection{The velocity switching model}
An interesting class of stochastic models, originally proposed to study turbulent motion was developed in \cite{Tay} and \cite{Gol}. These models capture the essence of a microscopic model of a the randomly moving particle. Such models were eventually extended to a larger class of models under the name of random evolutions see \cite{K1,Her,Pin,Swi,Kol}, or piecewise deterministic dynamical systems, \cite{Fle,Da}.  Such dynamical systems can be summarily described as consisting of a collection of evolution (semi)groups on some appropriate function space, plus a Markovian process on some auxiliary state space (aptly named ``the environment''), and a rule specifying which evolutions semigroup occurs in each possible environment. 
 
The model proposed in \cite{Gol} goes as follows: A particle starts at some point $x\in\bbR$ and moves to the right (or to the left) with velocity $v>0$ (or $-v$).  That is, ``the environment'' has two states modeled by an auxiliary two-state Markov chain. The chain alternates states at random times, which are exponentially distributed, and when the chain switches states, the particle changes the sign of its velocity. Let $\lambda$ denote the frequency of jumps, that is, its inverse $1/\lambda$ is the mean lifetime at each state, and for the time being we assume that it is the same for both states. When relativistic effects are of interest, we will assume that the transition rate is that measured by an observer at rest with respect to the environment.

The Markov property means that there is a sequence of independent, identically distributed random variables, denote them by $\{\tau_n: n=1,2,...\}$ denoting the times between velocity switches. Denote by $T_n=\sum_{j=1}^n\tau_j$ the (random) time it takes for the first $n$-velocity switches to occur, and we put $N(t)=n$ if $T_n\leq t <T_{n+1}$ to denote the number of transitions up to time $t.$ The process $N(t)$ is Poisson with parameter $\lambda t.$ In particular, if $t_1<t_2,$ then $N(t_1)$ and $N(t_2)-N(t_1)$ are independent, and $N(t_2)-N(t_1)$ has the same distribution as $N(t_2-t_1).$.   

 The trajectories of the velocity-switching process are easily established to be
\begin{equation}\label{mot1}
\begin{aligned}
\nu(t) = \nu(0)(-1)^{N(t)}\,\;\;\; & t \geq 0, \;\;\;\mbox{where}\;\;\;\nu(0)=\pm v.\\
x(t) = x& + \int_0^t \nu(s)ds.
\end{aligned}
\end{equation}
We use $\nu(t)$ instead of the more natural $v(t)$ to emphasize the random nature of the velocity process. It is well known that the joint process $(x(t),\nu(t))$ is a Markov process with state space $\bbR\times\{-v,v\}.$ Actually, it is the simplest case of a Markov additive process, (see \cite{Cin} for the basics about such a class). 

We are assuming the existence of an underlying probability space on which all random variables are defined, and let $P$ denote the probability with respect to which the process $(x(t),\nu(t))$ is Markov, and let $E$ denote the expectation with respect to $P.$ We will use the standard notations $P^{x(0),\nu(0)}(A)$ (resp. $P^{\nu})(A)$ to denote the conditional probabilities $P(A|x(0),\nu(0))$ (resp. $P(A|x(0),\nu(0))$) of an event $A$ given its initial position and initial velocity. Similarly, if $X$ is an integrable random variable, $E^{x(0),\nu(0)}[X]$ (resp. $E^{\nu}[X]$) to denote the conditional expected values $E[X|x(0),\nu(0)]$ (resp. $E[X|\nu(0)],$)

 The first thing that is clear from \eqref{mot1} is that, for any time $t>0$ the particle is located within $[x-vt,x+vt]$ with probability $1.$ Not only that, the trajectories are of bounded variation (rectifiable) and, in any time interval, they are differentiable except at a finite number of values of $t,$ corresponding to the times $T_n$ of occurrence of the switching times of the velocity.  So, this model is acceptable from the physical point of view.
 
The following result (proved in \cite{Pin,Kol} among many places) will be used below. If $F:\bbR\times\{-v,v\}\to \bbR$ is a function defined on  the state space of the process, and if we put $u(t,x,\pm v) = E^{x,\pm v}[F(x(t),\nu(t))],$ then $u(t,x,\pm v))$ solves the (backward) Chapman-Kolmogorov system:
\begin{equation}\label{syst0}
\begin{pmatrix}  
\frac{\partial}{\partial t}-v\frac{\partial}{\partial x} & 0\\
0  & \frac{\partial}{\partial t}+v\frac{\partial}{\partial x}
\end{pmatrix}{u(t,x,v)\atopwithdelims() u(t,x.-v)} =
\lambda\begin{pmatrix} -1 & 1\\ 1 & -1\end{pmatrix}{u(t,x,v)\atopwithdelims() u(t,x.-v)},
\end{equation}
with initial conditions $u(0,x,\pm v)=F(x,\pm v).$ 
To decouple this system, differentiate each one with respect to $t,$ using the system in the intermediate step, it follows that each of $u(t,x,\pm v)$ satisfies the telegraph equation:
\begin{equation}\label{syst0.1}
\frac{\partial^2}{\partial t^2}u + 2\lambda\frac{\partial}{\partial t}u = v^2\frac{\partial^2}{\partial x^2}u.
\end{equation}
The initial conditions for each are as above $u(0,x,\pm v)=F(x,\pm v),$ whereas $\partial u(0,x,\pm v)/\partial t$ are obtained from \eqref{syst0} evaluating at $t=0.$  Even though the equations are decoupled as equations, they are coupled through their initial conditions.

If we denote by $f_{+}(t,x)$ (respectively, by $f_{-}(t,x)$) the probability density of finding the particle in $(x,x+dx)$ at time $t>0,$ if it started with speed $v$ (respectively $-v$), then these quantities satisfy the (forward) Chapman-Kolmogorov system:
\begin{equation}\label{syst1}
\begin{pmatrix}  
\frac{\partial}{\partial t}+v\frac{\partial}{\partial x} & 0\\
0  & \frac{\partial}{\partial t}-v\frac{\partial}{\partial x}
\end{pmatrix}{f_{+}\atopwithdelims() f_{-}} =
\lambda\begin{pmatrix} -1 & 1\\ 1 & -1\end{pmatrix}{f_{+}\atopwithdelims() f_{-}}.
\end{equation}
As above, each component satisfies:
\begin{equation}\label{syst2}
\frac{\partial^2}{\partial t^2} f_{\pm} + 2\lambda\frac{\partial}{\partial t} f_{\pm} = v^2\frac{\partial^2}{\partial x^2} f_{\pm}.
\end{equation}
The equations can be integrated separately, and as for \eqref{syst1}, they are coupled through their initial conditions.

At this point, we mention another reason why the diffusion equation \eqref{eq1.1} is not compatible with special relativity. For proofs and earlier references see \cite{Pin} for example. Note that $1/\lambda$ is the mean time between shocks, when it goes to zero, that is, when $\lambda \to \infty,$ and $v\to \infty$ in such a way that $v^2/\lambda = \sigma$ is constant, then each \eqref{syst0.1} or \eqref{syst2} becomes the diffusion equation. This argument implies that $v\approx \sqrt{\sigma\lambda}$ becomes larger than the speed of light for large enough $\lambda$ and that, since the number of velocity switchings per unit time tends to infinity,  the trajectories of the limiting process, are not differentiable at infinitely many points in each time interval.  

\subsection{The contents of this paper and a look at the literature}
Having introduced the main notations, after describing the contents of the paper,  in the remainder of this section we review some of the large body of literature on this subject and its connection to Dirac's equation. The rest of the paper is organized as follows. In Section 2 we examine a simple Hamiltonian system that leads to \eqref{syst1}. One reason for that is that we want to cast this work in the framework of \cite{Gz}, where the notion of quantum random evolution was introduced for systems with a finite number of states. The other reason is that the classical Hamiltonian can be easily transformed into a quantum Hamiltonian. In Section 3 we consider the covariance of the system \eqref{syst1} relative to the Lorentz transformations. We prove that when the trajectories in each coordinate system are parameterized by their proper time, then each of \eqref{syst1} is covariant with respect to the Lorentz transformation. In Section 4 we deal with the quantum case, that is, we start with the classical Hamiltonian introduced in Section 2, and compute the time dynamics of the expected value of some observables.

In \cite{Tay,Gol,CV,K1} some probabilistic aspects of \eqref{mot1} thought of as a stochastic process were examined. In particular, in \cite{CV} the invariance of \eqref{syst2} with respect to changes of coordinates in space-time was discussed. There are several higher dimensional extensions of \eqref{mot1}. An interesting one comes under the name of {\it isotropic transport on manifolds}. See \cite{Pin,Kol} for details and reference to earlier work. See also \cite{Kol2} for more on the computation of higher dimensional transition probabilities. A different stochastic process leading to the telegraph equation and a different integration of it are presented in \cite{TS}. 

For other models that include special relativity and have the diffusion equation as the non-relativistic limit, see \cite{DH1,DH2} and the well-documented review \cite{DH3} 

 The formal similarity of the system \eqref{syst1} to Dirac's equation in $1+1$ dimensions has motivated much work on the probabilistic approach to solve Dirac's equation using path integration. Consider for example \cite{GJK,MO} as well as \cite{JS,JK} and for a connection to Maxwell's equations see \cite{Ord}. Related work on relativistic kinematics, the Poisson-Ka\c{c} model can be seen in \cite{Gio1,Gio2,Gio3,Gio4}. In \cite{BNO} \hl{the authors make use of a Lorentz transformation to eliminate the drift term in} \eqref{syst2}. For an interesting application to neuronal models see \cite{Rat}. And for applications in option pricing in finance, including a long list of relevant references, see \cite{RK}.
 
A related class of processes, that have probability densities satisfying equations like the telegraph equation, are the persistent random walk processes. For review and further references consider \cite{ML,MLW,Wei}.

The quantum random evolutions as used here, were introduced in \cite{Gz}. That work contains the basics for the extension of the theory of random evolution to the quantum mechanical setup, and the Hamiltonian framework for the velocity switching process considered here, was mentioned as a possible application. To finish we mention \cite{FH,BH,LYM,Zie}, that have names similar to \cite{Gz}, but which deal with different problems with vaguely related techniques, or that deal with similar problems but different techniques.

\section{The Hamiltonian approach to the random evolution}
The following pair of (parameterized) Hamiltonians, that yield the dynamics like that of the Goldstein-Ka\c{c} model, were proposed in \cite{Gz}. 

\begin{equation}\label{ham1}
\begin{aligned}
H_1 = vp\\
H_2 = -vp.
\end{aligned}
\end{equation}

Recall that we suppose that $v>0.$ The corresponding Hamilton equations of motion are:

\begin{equation}\label{mot3}
\begin{aligned}
\frac{dx}{dt} = \frac{\partial H_1}{\partial p} = v,\;\;\;\mbox{and}\;\;\; \frac{dp}{dt} = -\frac{\partial H_1}{\partial x} = 0.\\ 
\frac{dx}{dt} = \frac{\partial H_2}{\partial p} = -v.\;\;\;\;\mbox{and}\;\;\; \frac{dp}{dt} = -\frac{\partial H_2}{\partial x} = 0.
\end{aligned}
\end{equation}
Keep in mind that in this model the velocity $v$ is a fixed phenomenological parameter for each evolution. Even though the momentum is a constant of the motion, it is not a priori related to the velocity. It plays the role of the infinitesimal generator of displacements through the Lie algebra defined by the Poisson brackets. The Hamiltonians are singular. Their conjugate functions defined by the Legendre transformation $L(\dot{x}) =\sup\{p\dot{x}-H(p)\}$ are $\delta_{\pm v}(\dot{x})$-like function that have value $0$ when $\dot{x}=\pm v$ or $+\infty$ otherwise.

To continue, if we integrate the equations of motion, subject to initial conditions $x(0)=x_0$ and $p(0)=p_0$ we obtain:
\begin{equation}\label{em1}
x(t) = x_0 \pm vt\;\;\;\; \mbox{and}\;\;\; p(t)=p_0.
\end{equation}
These solution can be regarded from three points of view: \hl{First, as the trajectories of a dynamical system. Second as characteristics of the linear partial differential equation} \eqref{syst1} for example (see \cite{HS,J}). \hl{And third, from the point of view of Markov processes theory, which bridges the former two.}

This solution induce two flows in phase space. But since $p$ is a constant of the motion, we can concentrate on the flow in the coordinates only. Define the flows:
\begin{equation}\label{flow1}
\phi_{+}(x) = (x+vt),\;\;\;\;\mbox{and}\;\;\;\;\phi_{-}(x) = (x-vt)
\end{equation}

These induce the following transforms on the space of functions defined on the coordinate space. Given $F(x)$ we have: 
\begin{equation}\label{flow2}
\Phi_t^{+}F(x) = F(\phi_{+}(x) = F(x+vt),\;\;\;\;\mbox{and}\;\;\;\;\Phi_t^{-}F(x) = f(\phi_{-}(x)) = F(x-vt).
\end{equation}
To think of them as Markov processes, we write $X(t)=X(0)\pm vt.$ Note that
$$P(X(t) \in A|X(0)=x) = \int_A \delta\big(y - (x\pm vt)\big)dy$$
equals $1$ if $x\pm vt\in A$ and $0$ otherwise. This means that, say, 
$E[F(X(t))|X(0)=x]=\Phi_t^{\pm v}F(x).$ \hl{Please, note that we have two flows, one for each initial sign of the velocity}. The infinitesimal generators of these flows are

\begin{equation}\label{infflow1}
\frac{\partial\Phi_t^{\pm}F(x)}{\partial t}  = \pm v\frac{\partial\Phi_t^{\pm}F(x)}{\partial x}
\end{equation}
with initial conditions $\Phi_0^{\pm}F(x)=F(x).$ These are the backward Champan-Kolmogorov equations of these simple Markov processes. To bring in the transition densities, we write:
$$\Phi_t^{\pm}F(x) = E[F(X(t)|X(0)] = \int F(y)\rho_{\pm}(t,x,y)dy$$
where $\rho_{\pm}(t,x,y)=\delta\big(y-(x\pm vt)\big).$ The evolution equation for the transition density is obtained by differentiating both sides with respect to $t,$ making use of the explicit form of the density, we obtain: 
$$
\frac{\partial\rho_{\pm}}{\partial t}  = \mp v\frac{\partial\rho_{\pm}}{\partial x }.
$$
 Rearranging, we obtain the forward Chapman-Kolmogorov equation for the transition densities: 
\begin{equation}\label{infflow12}
\frac{\partial\rho_{\pm}(t,x,y)}{\partial t}  \pm v\frac{\partial\rho_{\pm}(t,x,y)}{\partial x } = 0.
\end{equation}
The trajectories in \eqref{flow1} are the characteristic curves of these linear PDE's. These are the homogeneous parts of \eqref{syst1}.\\  

\subsection{Intertwining the Hamiltonian flows to obtain the random evolution}
The two trivial Hamiltonians flows will now be intertwined (entangled) using a Markov chain (the environment) that flips the sign of the parameter $v$. Denote by  $\bz(t)=(x(t),\nu(t))$ the Markov process on $\bbR\times\{-v,v\}$ introduced in Section 1.1. Explicitly:
\begin{equation}\label{mkv}
\bz(t) = (x + \int_0^t\nu(s)ds, \nu(t))
\end{equation}
 It is continuous in the first component and left continuous in the second, and it is worth keeping in mind that $\nu(t)= \nu(T_{n-1})$ for $T_{n-1}\leq t<T_n.$ If we know that only $n$ shocks have occurred, we can write
\begin{equation}\label{mkv2}
\bz(t) =(x+\sum_{i=1}^n\nu(T_{i-1})\tau_i + \nu(T_n)(t-T_n), \nu(T_n))\;\;\;\mbox{when}\;\;\;T_n\leq t<T_{n+1}.
\end{equation}

The first component refers to the position, and the second is the velocity. The random evolution operator acting on $F(\bz)$ is defined by:

\begin{equation}\label{rev1}
\cT_t F(\bz) = \Phi_{t-T_n}^{\nu(t)}\Phi_{\tau_n}^{\nu(n-1)}...\Phi^{\nu(0)}_{\tau_0}F(\bz) = F(\bz(t)),\;\;\;\mbox{whenever}\;\;\;N(t)=n.
\end{equation}
\hl{As above, the superscripts stand for the velocity at the beginning of the corresponding time interval, the subscript indicates at which time the position is to be computed according to} \eqref{mkv2}. This is just a different way of referring to \eqref{mkv2}. This describes the random flow. In the classical world, if we want to make predictions, we have to compute

\begin{equation}\label{pred1}
\begin{aligned}
u(t,x,1) = E^{(x,+v)}[\cT_t F(\bz)] = E^{(x,+v)}[F(x(t),\xi(t))]. = \sum_{j=1}^2\int F(y,j)f_{1,j}(t,y)dy\\
u(t,x,2) = E^{(x,-v)}[\cT_t F(\bz)] = E^{(x,-v)}[F(x(t),\xi(t))] = \sum_{j=1}^2\int F(y,j)f_{2,j}(t,y)dy.
\end{aligned}
\end{equation}
For each $j=1,2,$ the densities $f_{i,j}(t,y)$ satisfy an equation like \eqref{syst1}. These are explicitly \hl{obtained} in \cite{Pin} for example. Instead, one could solve a system like \eqref{syst1} for the $u(t,x,i).$ They satisfy the backward Chapman-Kolmogorov system:

\begin{equation}\label{syst3}
\begin{pmatrix}  
\frac{\partial}{\partial t}-v\frac{\partial}{\partial x} & 0\\
0  & \frac{\partial}{\partial t}+v\frac{\partial}{\partial x}
\end{pmatrix}{u_{1}\atopwithdelims() u_{2}} =
\lambda\begin{pmatrix} -1 & 1\\ 1 & -1\end{pmatrix}{u_{1}\atopwithdelims() u_{2}}.\;\;\;\mbox{with}\;\;\;{u_{1}(0,x)\atopwithdelims() u_{2}(0,x)}={F(x,1)\atopwithdelims() F(x,2)}
\end{equation}

Before closing, to examine how a Hamiltonian like $H(x,p)=vp$ behaves relative to Galilean transformations, consider the generation function $G(x,P,t)=(x-Vt)P.$  This generates a change to a system moving with speed $V$ relative to the original system. The transformation equations are (see \cite{Gol2}):
$$X=\frac{\partial G}{\partial P} = x-Vt,\;\;\;p=\frac{\partial G}{\partial x}=P,\;\;\;\hat{H}(X,P)=vP+\frac{\partial G}{\partial t} = (v-V)P.$$
In the $(X,P)$ system, the velocity of the particle is $\dot{X}=v-V$ as it should be. This simple example suggests that if initially, the forward speed $v$ and the backward speed
$-v$ were not exact opposites, by jumping onto a moving system, one can see the process from a coordinate system in which they are equal with opposite signs.

\section{Covariance of the physical Brownian motion relative to Lorentz transformations}

In this section, we analyze the transformation properties of any of the equations in \eqref{syst1} with respect to a Lorentz transformation. Consider the generic equation:
\begin{equation}\label{test1}
\frac{\partial g}{\partial t} + v\frac{\partial g}{\partial x}=(\sqrt{1-(v/c)^2})\lambda h =\tilde{\lambda}h.
\end{equation}
To interpret the right hand side, think of a decay process, which for an observer in the system in which the mechanism is at rest, has frequency $\lambda.$ For an observer moving with speed $v,$ the frequency will be $\tilde{\lambda}=\sqrt{1-(v/c)^2}\lambda.$ The argument put forward for this case applies to the backward and the forward Chapman-Kolmogorov systems. That is why it suffices to consider just one of them. Later on, we shall explain why both equations of the system transform nicely.

We shall call the system in which the coordinates are $(t,x)$ the laboratory (or coordinate) frame   (or system). To take into account relativistic effects, we multiply the transition rate (which is the inverse of a decay time) by the time contraction factor in the laboratory system.

We want to examine how an observer in a frame that moves with constant speed $V,$ describes \eqref{test1}. Denote the coordinates of the observer in the moving frame by $(t',x').$ The space-time coordinate transformation between the two observers is
\begin{equation}\label{lor1}
{t' \atopwithdelims()x'} = \gamma\begin{pmatrix}
                                          1 &  -V/c^2\\
                                          -V& 1\end{pmatrix} {t \atopwithdelims()x}\;\;\;\mbox{and} \;\;\;
{t \atopwithdelims()x} = \gamma\begin{pmatrix}
                                          1 &  V/c^2\\
                                          V& 1\end{pmatrix} {t'\atopwithdelims()x'}.
\end{equation} 
As usual, $\gamma=1/\sqrt{1-(V/c)^2}.$ Besides, in the moving frame, the speed is  $v'= (v-V)/(1-vV/c^2).$ Write $(t',x')=L(t,x)$ and define $\tilde{g}(t',x')=g(L(t,x))$ and $\tilde{h}(t',x')=h(L(t,x)).$   In order to examine the covariance of \eqref{test1} we need to compute
$$\frac{\partial \tilde{g}}{\partial t'} + v'\frac{\partial \tilde{g}}{\partial x'}.$$
Recall that the characteristic curves for the differential operators, in each of the two coordinate systems are $(t,x-vt)$ and, respectively, $(t', x'-v't').$                            This means that the proper times along each of them satisfy:
\begin{equation}\label{proptime}
\frac{dt}{d\tau} = \frac{1}{\sqrt{1 - (\frac{v}{c})^2}}\;\;\;\;\;\;\frac{dt'}{d\tau'} = \frac{1}{\sqrt{1 - (\frac{v'}{c})^2}}.
\end{equation}
Using the transformation equations $(t',x')=L(t,x)$ in \eqref{lor1} and the chain rule, one can verify that
$$\frac{\partial \tilde{g}}{\partial t'} + v'\frac{\partial \tilde{g}}{\partial x'}=\frac{\sqrt{1-(V/c)^2}}{ \big(1-vV/c^2\big)}\bigg(\frac{\partial g}{\partial t} + v\frac{\partial g}{\partial x'}\bigg).   $$

An arithmetic manipulation (see \cite{MK} in case of need), transform the addition of velocities formula into:
$$\frac{1}{\sqrt{1 - (\frac{v'}{c})^2}} = \frac{\big(1-vV/c^2\big)}{\sqrt{1 - (\frac{V}{c})^2}\sqrt{1 - (\frac{v}{c})^2}}$$
 If we multiply the last identity by the former, and bring in the identities in \eqref{proptime}, we obtain:
 $$
 \frac{\partial \tilde{g}}{\partial \tau'} + \frac{dx'}{d\tau'}\frac{\partial \tilde{g}}{\partial x'}= \bigg( \frac{\partial g}{\partial \tau} + \frac{dx}{d\tau}\frac{\partial g}{\partial x}\bigg).$$
 Now, use \eqref{test1} to obtain that in the coordinates of the moving system the transport equation is

 \begin{equation}\label{test2}
\frac{\partial \tilde{g}}{\partial t'} + \frac{dx'}{dt'}\frac{\partial \tilde{g}}{\partial x'}
=\sqrt{1-(v'/c)2}\lambda\tilde{h}.
\end{equation}
This is the generic evolution equation in the $(t',x')$ coordinate system equivalent to \eqref{test1}. Therefore, \eqref{test1} transforms covariantly. 

We already mentioned that from the definition of the Lorentz transformation it is easy to see that the velocities $v$ and $-v$ in the moving frame become:
$$v'= \frac{v-V}{1-\frac{vV}{c^2}}\;\;\;\;\mbox{and}\;\;\;\;v'' = -\frac{v+V}{1+\frac{vV}{c^2}}$$

We can now apply the previous computations to $\tilde{f}_{+}(t',x')=f_{+}(L(t,x))$ and $\tilde{f}_{-}(t',x')=f_{-}(L(t,x)),$ and repeat the above computations once with $v'$ and once with $v'',$ to obtain t  

The above computations imply that under a Lorentz transformation, in the coordinates used by the moving observer, the system \eqref{syst1}, becomes:

\begin{equation}\label{syst2.0}
 \begin{pmatrix}  \frac{\partial}{\partial t'}+v'\frac{\partial}{\partial x'} & 0\\
                              0  & \frac{\partial}{\partial t'}+v''\frac{\partial}{\partial x'}
                         \end{pmatrix} {\tilde{f}_{+}\atopwithdelims() \tilde{f}_{-}} =
\\ \begin{pmatrix} -\tilde{\lambda}' & \tilde{\lambda}'\\ 
                                \tilde{\lambda}'' & -\tilde{\lambda}''
                           \end{pmatrix}{\tilde{f}_{+}\atopwithdelims() \tilde{f}_{-}}.
\end{equation}
We put $\tilde{\lambda}'=\sqrt{1-(v'/c)^2}\lambda$ and $\tilde{\lambda}''=\sqrt{1-(v''/c)^2}\lambda.$  \hl{In other words, the system transforms properly under Lorentz transformation if the transition rates are interpreted as transition rates seen by the moving particle for each observer. In other words, the Chapman-Kolmogorov system describing the random evolution of the is Lorentz covariant when the change of the transition rates with respect to the Lorentz transformation is taken into account}.

\section{The quantum case}
Here we develop the quantum version of the Goldstein-Ka\c{c} model. For that, we quantize the classical Hamiltonians proposed in Section 2, then, we find out the unitary time evolution operators that they define on the states space of the system, and then we show how these time evolutions can be intertwined to obtain the quantum Brownian motion.
As said, the time evolution of the system is governed by the Hamiltonians:
\begin{equation}\label{qhams1}
\hat{H}_{\pm} = \pm v\hat{p}.
\end{equation}
Recall that $v>0$ is a phenomenological speed that characterizes each particular system.  We use the conventional notation $\hat{p}$ to emphasize the operator nature of the momentum.
 If we stay in the Schr\"{o}dinger picture, and use the Bohr-Sommerfeld correspondence rules, in the absence of shocks, the time evolution equations that the two Hamiltonians imply are:

\begin{equation}\label{qte2}
i\hbar\frac{\partial}{\partial t} \psi_{\pm} = \hat{H}\psi_{\pm}= -i\big({\pm}v\big)\hbar\frac{\partial}{\partial x}\psi_{\pm}\;\;\;\;\Leftrightarrow\;\;\;\; \frac{\partial}{\partial t} \psi_{\pm} = \mp v\frac{\partial}{\partial x} \psi_{\pm}.
\end{equation}

In other words, for any initial state $\psi_{0,\pm}(x),$ in the coordinate representation, the time evolution that each of the Hamiltonians induces is:
\begin{equation}\label{te1}
\psi_{\pm}(t,x) = U_{\pm}(t)\psi_{\pm}(x) = e^{-it\hat{H}_{\pm}/\hbar}\psi_{\pm}(x) = e^{\mp tv}\psi_{\pm}(x) = \psi_{0,\pm}(x  \mp vt).
\end{equation}
Except for the interpretation of the $\psi_{\pm},$ these coincide with those of the probability densities of the trivial Markov process on the configuration space.  

To obtain the quantum random evolution, we intertwine these unitary time evolutions according to the velocity-switching mechanism. \hl{We shall denote the stated of the environment at time} $t$ by $\xi(t),$ and we will shift between $\pm$ and $1,2$ as subscripts denoting the random state of the environment whenever it is typographically more convenient. \hl{Keep in mind that now $v>0$ is a phenomenological parameter}. If the current time is $t,$ and $N(t)$ shocks have occurred. the states of the random environment are $\xi(0),\xi(1),...,\xi(T_{N(t)}).$ Then the time evolution operators during the intervals $[0,T_1),[T_1,T_2),...[T(N(t)),t-T(N(t)))$ are respectively $U_{\xi(0)}(T_1), U_{\xi(T_1)}(T_2-T_1),..,U_{\xi(0)}(T(N(t))(t-T(N(t))).$

 Then, in the coordinate representation, if the initial state is $\psi_{\xi(0)}(x),$ the state at time $t$ is:
\begin{equation}\label{qre1}
\cU_{\xi(0)}(t)\psi_{\xi(0)}(x) = U_{\xi(T_{N(t)})}(t-T_{N(t)})U_{\xi(T_{N(t)-1})}(T_{N(t)}-T_{N(t)-1})...U_{\xi(0)}(T_1)\psi_{\xi(0)}(x).
\end{equation}
Invoking the commutativity of these evolutions and \eqref{te1}, we can write this as:
\begin{equation}\label{qre2}
\psi_{\xi(t)}(t,x) = \cU_{\xi(0)}(t)\psi_{\xi(0)}(x) = \psi_{\xi(0)}(x-v\int_0^t\xi(s)ds).
\end{equation}
That is, for each possible random sequence of environments, we have one possible state at time $t.$ Therefore prediction involves two stages: first, the quantum expectation for each given history of the environment, and second, the statistical average of the quantum expected value over all the histories of the environment.

Let $f(\hat{x})$ be a function of the position operator. Its quantum expected value at time $t,$ when the system is in the random state $\psi_{\xi(t)}(t,x)$ is;

\begin{equation}\label{ave1}
\langle\psi_{\xi(t)}(t,x),f(\hat{x})\psi_{\xi(t)}(t,x)\rangle = \int f(x)|\psi_{\xi(t)}|^2(t,x)dx.
\end{equation}
It is here where the simple nature of the evolution comes in. According to \eqref{qre2}: 
$$|\psi_{\xi(t)}|(t,x)^2 =| \psi_{\xi(0)}|^2\big(x-v\int_0^t\xi(s)ds\big).$$
That is, for each sequence of random environments, the quantum expected value of 
$f(\hat{x})$ in the random state $\psi_{\xi(t)}(t,x)$ is explicitly given by:
\begin{equation}\label{ave2}
\langle\psi_{\xi(t)}(t,x),f(\hat{x})\psi_{\xi(t)}(t,x)\rangle = \int f(x)| \psi_{\xi(0)}|^2\big(x-\int_0^t \xi(s)ds\big)dx.
\end{equation}

And now, we have to average over all possible environments, or, over all the trajectories of the underlying Markov chain up to time $t.$ Let us write this as:

\begin{equation}\label{ave3}
\begin{aligned}
\langle f(\hat{x})\rangle_{ave}(t) =& E^{\xi(0)}[\langle\psi_{\xi(t)}(t,x),f(x)\psi_{\xi(t)}(t,x)\rangle]\\ = E^{\xi(0)}[\int f(x)| \psi_{\xi(0)}|^2\big(x-\int_0^t \xi(s)ds\big)dx = &\int f(x)\bigg( E^{\xi(0)}[|\psi_{\xi(0)}|^2\big(x-\int_0^t \xi(s)ds\big)]\bigg)dx. \\
\end{aligned}
\end{equation}

Note that, since there are two possible signs for $\xi(0),$ we have two different values for the function inside integral, that is:
$$\rho^{ \pm}(t,x) = E^{(x,\pm v)}[|\psi_{\xi(0)}|^2\big(x(0)+v\int_0^t\xi(s)ds\big))].$$
With these notations, we now write \eqref{ave3} as:

\begin{equation}\label{ave4}
\langle f(\hat{x})\rangle_{ave}(t) = \int f(x)\rho^{\pm}(t,x) dx.
\end{equation}

As mentioned above, the vector valued function $(\rho^{+}(t,x),\rho^{-}(t,x))^t$ satisfies a system like \eqref{syst0} or \eqref{syst3}, namely:

\begin{equation}\label{ave3.1}
\begin{pmatrix}  
\frac{\partial}{\partial t}+v\frac{\partial}{\partial x} & 0\\
0  & \frac{\partial}{\partial t}-v\frac{\partial}{\partial x}
\end{pmatrix}{\rho^{+}\atopwithdelims() \rho^{-}} =
\lambda\begin{pmatrix} -1 & 1\\ 1 & -1\end{pmatrix}{\rho^{+}\atopwithdelims() \rho^{-}}.
\end{equation}
We have already mentioned that the solution to this system can be computed explicitly. See \cite{Pin,Kol} for example. 
And in the previous section we saw that in the relativistic case, we should replace $\lambda$ by $\tilde{\lambda}=\big(1-v/c)^2\big)^{1/2}\lambda$ (the decay rate in the system in which the classical particle moves with speed $v$). In this case, the system is invariant under Lorentz transformations.

To continue with the quantum case: there is no loss of generality if we suppose that the initial condition for both is the same and equal to $|\psi_0(x)|^2.$  Notice that this is the relevant quantity to study, and not the average probability amplitude $E^{\xi(0)}[\psi(t,x)]$ which satisfies a similar system. To continue, each component of $\rho^{\pm}(t,x)$ satisfies \eqref{syst2} (or (\eqref{syst0.1}), that is:

\begin{equation}\label{syst5}
\frac{\partial^2}{\partial t^2}\rho^{\pm} + 2\lambda\frac{\partial}{\partial t} \rho^{\pm} = v^2\frac{\partial^2}{\partial x^2} \rho^{\pm}.
\end{equation}
If one applies the differential operator on the left-hand side of \eqref{syst5} to the left-hand side of \eqref{ave4}, one obtains

\begin{equation*}
\frac{\partial^2}{\partial t^2}\langle f(\hat{x})\rangle_{ave}(t) + 2\lambda\frac{\partial}{\partial t} \langle f(\hat{x})\rangle_{ave}(t) = \int f(x)\bigg(v^2\frac{\partial^2}{\partial x^2} \rho^{\pm}(t,x)\bigg)dx.
\end{equation*}
After integration by parts, we obtain:

\begin{equation}\label{ave5}
\frac{\partial^2}{\partial t^2}\langle f(\hat{x})\rangle_{ave}(t) + 2\lambda\frac{\partial}{\partial t} \langle f(\hat{x})\rangle_{ave}(t) = \int \rho^{\pm}(t,x)\bigg(v^2\frac{\partial^2}{\partial x^2} f(x)\bigg)dx.
\end{equation}

We will use \eqref{ave5} to compute the average position $\langle\hat{x}\rangle_{ave}(t) $ and the variance of the particle (when the initial state is $\psi_0(x)$). Considering first $f(x)=x$ and then, $f(x)=x^2,$ we obtain:
\begin{equation}\label{aux1}
\frac{\partial^2}{\partial t^2}\langle \hat{x}\rangle_{ave}(t) + 2\lambda\frac{\partial}{\partial t} \langle \hat{x}\rangle_{ave}(t) =  0,
\end{equation}
and:
\begin{equation}\label{aux2}
\frac{\partial^2}{\partial t^2}\langle \hat{x}^2\rangle_{ave}(t) + 2\lambda\frac{\partial}{\partial t} \langle \hat{x}^2\rangle_{ave}(t) =  2v^2.
\end{equation}
We solve the two equations, and in the appendix we explain how to obtain the integration constants. The first identity implies that:

$$\frac{\partial}{\partial t}\langle \hat{x}\rangle_{ave}(t) + 2\lambda \langle \hat{x}\rangle_{ave}(t) =  C_1= v+2\lambda\langle \hat{x}\rangle_{ave}(0)$$
Furthermore, $\langle \hat{x}\rangle_{ave}(0)=\langle \hat{x}\rangle_{\psi}(0)=\int x|\psi_0(x)|^2dx$ is the quantum expected value of $\hat{x}$ in the state $\psi(0).$  Integrating once more and rearranging we obtain:
\begin{equation}\label{mean}
\langle \hat{x}\rangle_{ave}(t) = \langle \hat{x}\rangle_{ave}(0)  + \frac{v}{2\lambda}\big(1 - e^{-2\lambda t}\big).
\end{equation}
 
 The average of the quantum variance over all random paths is
 $$\langle\Delta x\rangle_{ave}^2(t) \equiv  E^{\xi(0)}[\langle\psi(t),\big(\hat{x}-\langle \hat{x}\rangle_{ave}(t)\big)^2\psi\rangle] = \langle \hat{x}^2\rangle_{ave}(t) - \big(\langle \hat{x}\rangle_{ave}(t) \big)^2.$$ 

The identity \eqref{aux2} implies that
$$\frac{\partial}{\partial t}\langle \hat{x}^2\rangle_{ave}(t) + 2\lambda \langle \hat{x}^2\rangle_{ave}(t) = 2v^2t + C_2.$$
 In the appendix we verify that the integration constant is $C_2=2\lambda\langle\hat{x^2}\rangle_{\psi}(0)-v\langle \hat{x}\rangle_{\psi}(0).$
 Integrating once more we obtain:
 $$\langle \hat{x}^2\rangle_{ave}(t) =\langle \hat{x}^2\rangle_{\psi}(0)e^{-2\lambda t}+\frac{1}{\lambda}\big(\lambda\langle \hat{x}^2\rangle_{\psi}(0)-v\langle \hat{x}\rangle_{\psi}(0)\big)\big(1-e^{-2\lambda t}\big) +\frac{v^2}{\lambda}\big(1-e^{-2\lambda t}\big) -\frac{v^2}{2\lambda^2} \big(1-e^{-2\lambda t}\big).$$
Let us denote the standard deviation of the position operator by

$$\langle\Delta x\rangle_{ave}(t) \equiv \bigg((\hat{x}^2\rangle_{\psi}(t)-\big(\hat{x}\rangle_{\psi}(t)\big)^2\bigg)^{1/2}$$

Then, making use of \eqref{mean} to subtract $ \big(\langle \hat{x}\rangle_{ave}(t) \big)^2.$ using the previous result, canceling and rearranging we obtain

\begin{equation}\label{stdev}
\langle\Delta x\rangle_{ave}(t) = \bigg(\langle \Delta x\rangle^2_{\psi}(0) + \frac{v^2 t}{\lambda}\big(1-e^{-2\lambda t}\big) + R(\lambda,t)\bigg)^{1/2}.
\end{equation}
We put $(\langle \Delta x\rangle^2_{\psi}(0)=\langle\psi_0(\hat{x}-\langle\hat{x}\rangle_{\psi}(0)\big)^2\psi\rangle,$ where  $R(\lambda,t)$ is a bounded function of $t$ divided by $\lambda^2.$
Therefore, as $v\to \infty$ and $\lambda\to\infty$ in such a way that $\sigma=v^2/\lambda$ remains constant, from \eqref{stdev} we have:

\begin{equation}\label{asymp2}
\langle\Delta x\rangle_{ave}(t) \curly \bigg(\langle \Delta x\rangle^2_{\psi}(0) + \sigma t \bigg)^{1/2}.
\end{equation}
In other words, the average quantum variance is that of the initial quantum state corrected by the term similar to that of the classical Brownian motion. It is at this point where the incompatibility with special relativity creeps in: namely, for the asymptotics one requires $\sigma=v^2/\lambda$ to be constant while $v$ and $\lambda$ increase. This means the underlying microscopic motion must occur at speeds larger than $c.$

Notice that, the path averaged transition matrix satisfies (backward) Chapman-Kolmogorov system \eqref{ave3.1}, and we have already proved in Section 3, then this system is Lorentz invariant when the proper time is used in two coordinate systems moving with constant velocity relative to each other, and the system transforms properly when the average time ($1/\lambda$) is relativistically rescaled.

To finish, let us verify that the quantum version of the velocity switching process is compatible with special relativity. For that, suppose that $\psi_0(x)$ is such that $|\psi_0(x)|^2>0$ only for $x\in(a,b).$ In mathematical jargon, the support of $\psi_0$ is contained in $(a,b).$ According to \eqref{te1}, the state $\psi_{\pm}(t,x)$ at any $t>0,$ is such that $|\psi_{\pm}|^2(t,x)>0$ only in $(a-vt,b+vt).$  If $B=(c,d)$ is an interval such that $d<a-vt$ or $b+vt<c,$ or if $B$ does not intersect the support of $\psi_{\pm}(t),$ then the probability of finding the particle in $B$ is $0.$ To see that, put  $\psi_B(x)=\bbI_B(x)/\sqrt{(b-c)},$  which satisfies $\langle\Psi_B,\Psi_B\rangle=1.$ Therefore
$$\int_B |\psi_{\pm}|^2(t,x) = \tr\big(|\psi_B\rangle\langle\psi_B|\psi_{\pm}(t)\rangle\langle\psi_{\pm}(t)|\big)=0.$$ 
Therefore, if we average over all trajectories, we still have:
$$E^{\xi(0)}[|\langle\psi_B,\psi(t,x)\rangle|^2 ]= E^{\xi(0)}\big[ \int |\psi_B(x)|^2|\psi_0(x-\int_0^t\xi(s)ds)|^2dx \big]= 0.$$
In other words, the quantized version of the Goldstein-Ka\c{c} model, is consistent with special relativity in that no speed of propagation larger than $c$ are allowed.

\section{Closing remarks}
One could think of the following way of avoiding the unconventional Hamiltonian ($H=\pm vp$). Suppose that the Hamiltonian is that of the free particle ($H=p^2/2m$ say), and that the momentum reversals occur when the particle collides with supermassive particles separated by exponentially distributed distances along its path. Even though such a model may take care of the velocity reversals, it does not avoid the incompatibility with special relativity, for the simple reason that the classical momentum can in principle assume any value, and when quantized, the model leads to the usual quantum-free particle which is incompatible with the special relativity theory.

 What this implies is that to be consistent with special relativity, one has to work with Dirac's formulation of quantum mechanics from the outset.
 
 Some possible directions to explore, are the Hamiltonian version of \eqref{ham1} in which $v$ is random with a range contained in $(-c,c),$ or in a three-dimensional sphere of radius $c.$ This type of extensions have already been considered in probabilistic models known as random flights, see \cite{Pin,Kol}, for example. However, the extension of our presentation to the higher dimensional case does not lead to nice simple computations as in the one-dimensional case.

To finish, we mention \cite{CKB} in which the authors specify rules that coarse grained evolution equations have to satisfy to be consistent with Galilean transformations.

\section{Appendix: Pending computations}
Here we provide the determination of the constants $C_1$ and $C_2$ that appear in the integrations of \eqref{aux1}-\eqref{aux2}. Let us consider the first of the two equations in \eqref{ave3.1} and recall that for notational simplicity, we are supposing that $\psi_0(x)$ is independent of $v.$ Multiply the equation for $\rho_{+}(t,x)$ by $x$  and integrate over the real line. In the first term, exchange the time derivative with the integration, in the second integrate by parts, evaluate at $t=0$ to obtain:
$$\frac{\partial}{\partial t}\int x\rho_{+}(t,x)-v\int \rho_{+}(t,x) = 0.$$
The right hand is set equal to $0$ since $\int x\rho_{+}(0,x)=\int x\rho_{-}(0,x)$ by assumption.

Therefore, since $\int \rho_{+}(t,x)=1,$ it follows that $\frac{\partial}{\partial t}\int x\rho_{+}(t,x)=v\int \rho_{+}(t,x),$  and that $C_1=2\lambda\langle \hat{x}\rangle_{ave}(0)+v$ as claimed.
To obtain the constant $C_2$ for the integration of \eqref{aux2}, we proceed as in the previous case, except that now we multiply both sides of the first equation in the system \eqref{ave3.1} by $x^2.$ As above, the right-hand side vanishes at $t=0$ due to our assumption. Also, on the left-hand side, exchanging the time derivative with the integral of the first term, and integrating by parts of the second term, we are left with
$$\frac{\partial}{\partial t}\int x^2\rho_{+}(t,x)-v\int x\rho_{+}(t,x) = 0,$$
or in other words
$$\frac{\partial}{\partial t}\langle \hat{x}^2\rangle_{ave}(0)=2v\langle \hat{x}\rangle_{ave}(0)$$
Therefore $C_2=2\lambda\langle\hat{x}^2\rangle_{\psi}(0)+v\langle \hat{x}\rangle_{\psi}(0)$ as claimed.

\textbf{Data availability} There is no data availability issue associated with this submission.


\end{document}